\documentclass{svproc}
\usepackage{url}
\usepackage{amsmath}			
\usepackage{amssymb}			
\usepackage{color}				
\usepackage{esdiff}				
\usepackage{nicefrac}			
\usepackage[pdftex]{graphicx} 
\usepackage{here}			

\begin{document}
\mainmatter              
\title{The trouble with $2^{nd}$ order models or how to generate stop-and-go
traffic in a $1^{st}$ order model }
\titlerunning{Stop-and-go dynamics in a first order model}
%
\author{Jakob Cordes\inst{1} \and Andreas Schadschneider\inst{1} \and
Antoine Tordeux \inst{2}}
\authorrunning{J. Cordes, A. Schadschneider, A. Tordeux}
%
\tocauthor{Jakob Cordes and Andreas Schadschneider and
Antoine Tordeux}
\institute{Institut f\"ur Theoretische Physik,
  Universit\"at zu K\"oln,
  Germany\\
\email{jcordes@thp.uni-koeln.de, as@thp.uni-koeln.de}\\ 
\and
Institut f\"ur Sicherheitstechnik, Bergische Universit\"at
  Wuppertal, Germany
\email{tordeux@uni-wuppertal.de}
}
\maketitle              

\begin{abstract}
  Classical second order models of pedestrian dynamics, like the
  social-force model, suffer from various unrealistic behaviors in the
  dynamics, e.g. backward motion, oscillations and overlapping of
  pedestrians. These effects are not related to the discretization of
  the equations of motion, but intrinsic to the dynamics. They are the
  consequence of strong inertia effects that usually appear in second
  order models. We show that the experimentally
  observed stop-and-go behavior, which is an important test for any
  pedestrian model, can be reproduced with a stochastic first order
  model that does not suffer from the dynamical artefacts resulting
  from strong inertia. The model provides a new mechanism for
  stop-and-go behavior which is based on correlated noise.
\keywords{Pedestrian single-file motion, stop-and-go dynamics,
  first order microscopic models, coloured noise}
\end{abstract}
\section{Introduction}
In recent years, growing evidence suggests that
models of pedestrian dynamics based on second order differential
equations suffer from intrinsic problems that are not related to
numerical issues, e.g.\ insufficient discretization of the
differential equations.  In many second order models, unrealistic
behaviors like unmotivated backward motion, oscillations of the
direction of motion or overlapping for pedestrian and other traffic
models have been observed \cite{DAVIS2003,Wilson2004,Helbing2009}.

In \cite{Chraibietal}, the influence of the type of interaction force
on the dynamics has been studied in a one-dimensional single-file
scenario. It has been found that unrealistic behavior
  is related to instability phenomena and stop-and-go waves in these
models.  However, since stop-and-go waves have been observed in
experiments with pedestrians \cite{Portz,Zhangetal,Boltes2018} their
reproduction is a benchmark test for any model of pedestrian dynamics.

Here, we will address this problem from a slightly different
perspective. Instead of considering the influence of the specific
force used, we will argue that the inertia term in the second order
model is responsible for many unrealistic behaviors observed.
We propose a minimal first order model, i.e. a model which has no
inertia term related to the physical mass, which is able to
reproduce the basic properties of stop-and-go waves observed in
single-file experiments with pedestrians. The mechanism for the
formation of these waves is different than the classical mechanism
which is based on the fact that the homogeneous solution becomes
unstable. In our model, it is stable for all densities and the
stop-and-go waves are triggered by an additive {\em correlated}
noise.

\section{The Model}

The Optimal Velocity (OV) models are a class of models set by an
optimal velocity function $V(\cdot)$ which typically depends on the headway
$\Delta x$. The simplest OV model is the first order model
\begin{equation}
\dot{x}_n(t)=V(\Delta x _n (t)).
\end{equation}
In order to incorporate more realistic behaviour a reaction time
$\tau _R$ can be added as a delay \cite{Newell}
\begin{equation}
\dot{x}_n(t+\tau _R)=V(\Delta x _n (t)).
\label{OV_delayed}
\end{equation}
A Taylor expansion on the left hand side gives the second order model
proposed by Bando et al. \cite{Bando} with the sensitiviy
$ a = \frac{1}{\tau _R}$ .
Another important process in traffic is anticipation. Agents extrapolate
the current situation in order to reduce the reaction time
\cite{stoch_trans}. This helps to avoid collisions and allows a
smoother and faster flow.  Therefore an anticipation time $\tau _A$ is
added on the right hand side
\begin{equation}
  \dot{x}_n(t+\tau _R)=V(\Delta x _n (t+\tau _A)).
    \label {OVmodel}
\end{equation}
Obviously (\ref{OVmodel}) can be brought into the form of
(\ref{OV_delayed}) with $\tau_R \to \tau_R-\tau_A$ by
shifting the time. A specific form of the
Full-Velocity-Difference (FVD) model, which has been proposed in
\cite{Jiang}, can be derived when Taylor expansions are performed
independently in $\tau_A$ and $\tau_R$:
\begin{equation}
  \ddot{x} _n (t) = \frac{1}{\tau _R} \Bigl[V(\Delta x _n
  (t))-\dot{x}_n(t)\Bigr]
  + \frac{\tau _A}{\tau _R}\Delta \dot{x}_n (t) V'(\Delta x_n (t)).
  \label{FVDmodel}
\end{equation}
This is a second order model which does not have an inertia term
related to the physical mass $m$. Instead it has an effective
inertia which is determined by the two times $\tau_A$ and $\tau_R$.
The models (\ref{OVmodel}) and (\ref{FVDmodel}) have the same
stability condition
\begin{equation}
  \tau _R - \tau _A < \frac{T}{2},
  \label {stabcond}
\end{equation}
where a linear OV function $ V(d)=(d-l)/T $ with a desired time gap $T$
and a size $l$ of the agents has been assumed for congested states 
\cite{Jiang,TordeuxRL}. The case $\tau _A > \tau _R $ is unrealistic
because it corresponds to a motion which chooses the velocity optimal
according to the situation in the future. $\tau _R = \tau _A$
corresponds to a full compensation of the reaction time. If, in this
 case, (\ref{FVDmodel}) is combined with a white noise $\alpha \xi _n(t)$
  one gets
\begin{equation}
\begin{split}
\dot{x}_n (t)= V(\Delta x_n(t))+\epsilon _n (t); \\\
\dot{\epsilon} _n (t) =-\frac{1}{\tau} \epsilon _n (t) + \alpha \xi _n(t),
\end{split}
\label{modeldef}
\end{equation}
where the model has been rewritten as a first order OV model with a
correlated truncated Brownian noise $\epsilon_n(t)$ described by the
Orstein-Uhlenbeck process \cite{OU}. Eq.~(\ref{modeldef}) is the form
in which the model has been proposed in \cite{TordeuxS16}. In
accordance with condition (\ref{stabcond}) it is always linearly
stochastically stable, since the deterministic model is intrinsically
stable while the noise is additive and independent of the vehicle
states.  Note that (\ref{modeldef}) is a genuine first order equation
in the position variables $x_n$ since the second equation is the
definition of the noise that does not involve the variables $x_n(t)$.

\section{Simulations}

\begin{figure}[t]
	\centering
	\includegraphics[width=0.9\linewidth]{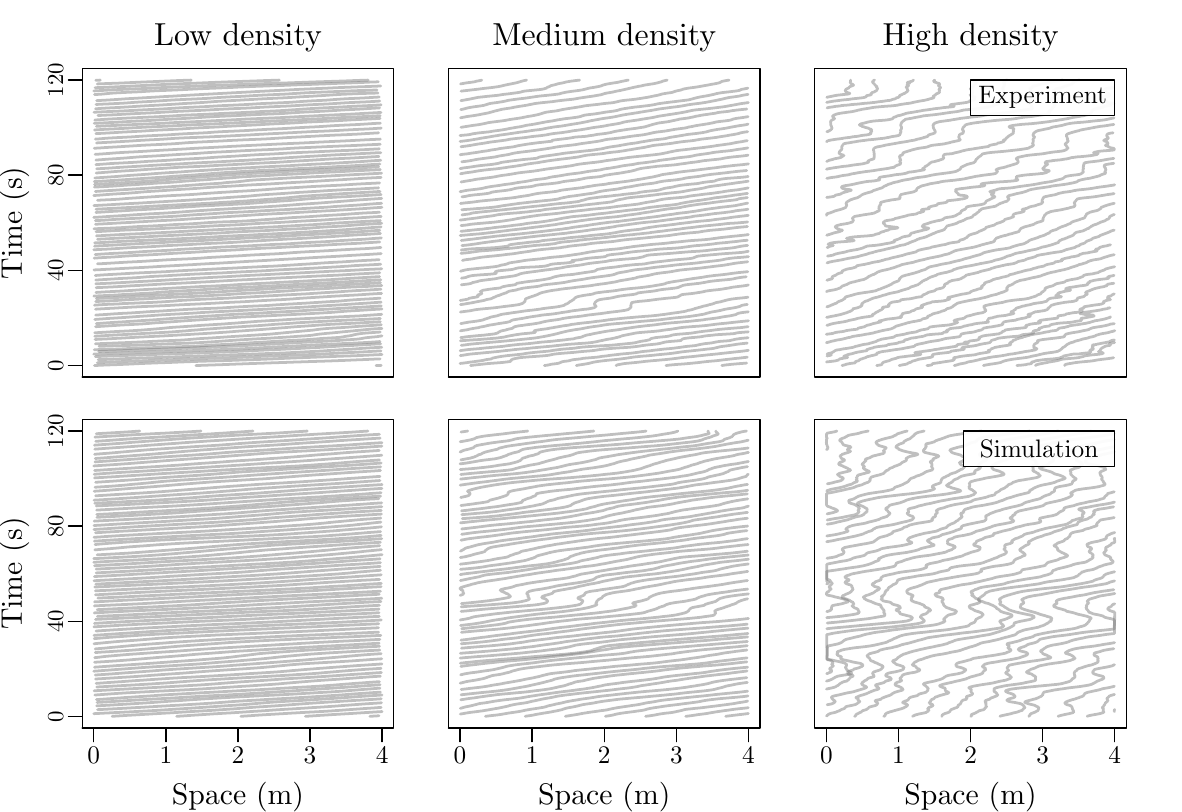}
	\caption{Empirical (top panels) and simulated (bottom panels)
          trajectories for different densities ($N=28, 45, 62$ from
          left to right). The initial configuration is homogeneous,
          both in the experiments and the simulations.}
	\label{fig:1}
\end{figure}

The model (\ref{modeldef}) is analyzed numerically using the explicit
Euler-Maruyama scheme with a time step $\delta t=0.01$~s. The
parameter values are $T=1.02$~s, $l=0.34$~m, $a=0.09$~ms$^{-3/2}$ and
$\tau = 4.4$~s, which are the statistical estimates in
\cite{TordeuxS16} with the data of \cite{Portz,Database}. The linear
OV function stated above is used. According to the corresponding
experimental situation the length is $L=27$m and the boundary
conditions are periodic.  Simulations are carried out with
$N=28, 45, 62$ pedestrians. For small densities a homogeneous free
flow state is observed ($N=28$), while stop-and-go waves appear at
higher densities ($N=45$ and $N=62$). The comparison with the
empirical trajectories shows good agreement (Fig.~\ref{fig:1}).

\begin{figure}[t]
	\centering
	\includegraphics[width=0.85\linewidth]{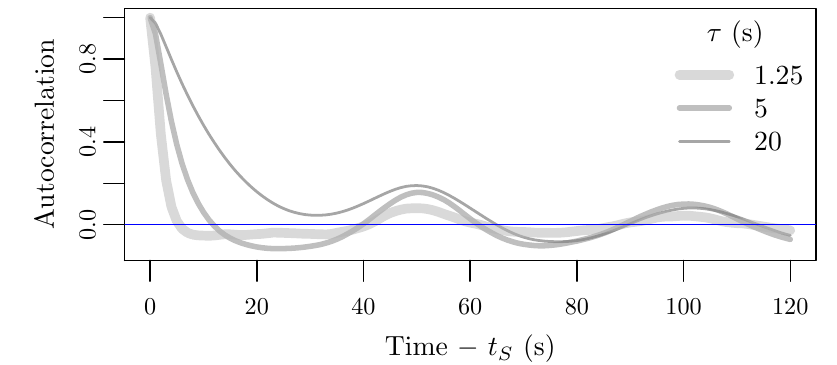}
	\caption{Mean temporal correlation function of the distance
          spacing in the stationary state of (\ref{modeldef}) for
          different values of the noise parameter $\tau$. Parameter
          $\alpha$ is chosen such that the noise amplitude is the same
          in all cases. The noise parameters do not influence the
          frequency of the waves, which only depends on $N$ and $T$.}
	\label{fig:2}
\end{figure}

The autocorrelation of the spacing for $50$ agents for different noise
parameters $\tau$ and $\alpha$ but a constant variance
$\sigma=\alpha\sqrt{\tau/2}$ is shown in Fig.~\ref{fig:2}. The system
can be supposed to be in the stationary state (simulation time
$t>t_s=2\cdot10^5s$).  The period of the autocorrelation remains the
same and stop-and-go behaviour is maintained in all cases but is less
pronounced for smaller $\tau$, i.e. less correlated noise. For a more
detailed analysis of the model we refer to \cite{TordeuxS16}.

The point to stress is the different mechanism of reproducing
stop-and-go behaviour. In most continuous models (e.g. social-force
model, deterministic OV models) the formation of a heterogeneous
configuration is the consequence of a deterministic instability and
can be described as a phase transition. In contrast, in model
(\ref{modeldef}) the stop-and-go waves are part of the stationary
state of the stochastic system. The system is permanently kicked out
of the deterministically stable homogeneous configuration by the
noise.  This mechanism can also be reproduced in the delayed first
order OV model, second order OV model and the FVD model if a white
noise is added in the deterministic stable regime. This can be
understood in the light of the connections between the models
investigated in the last section.

To explore this mechanism further, the effect of noise on a different
traffic model, namely the Gipps model, is investigated. In contrast
to the other models presented here, Gipps' model is time-discrete and
mainly considered for vehicular traffic. The update time $\tau$ is
also interpreted as the reaction time. It has a maximum acceleration
$a$, maximum deceleration $b$ and an estimated braking capability of
the preceding car $\hat{b}$, according to these the cars choose their
velocity. The stability of the homogeneous configuration of cars has
been investigated by Wilson \cite{wilson} and strongly depends on the
underestimation of the braking ability of the preceding car $\hat{b}$.

\begin{figure}[t]
	\centering
	\includegraphics[width=0.9\linewidth]{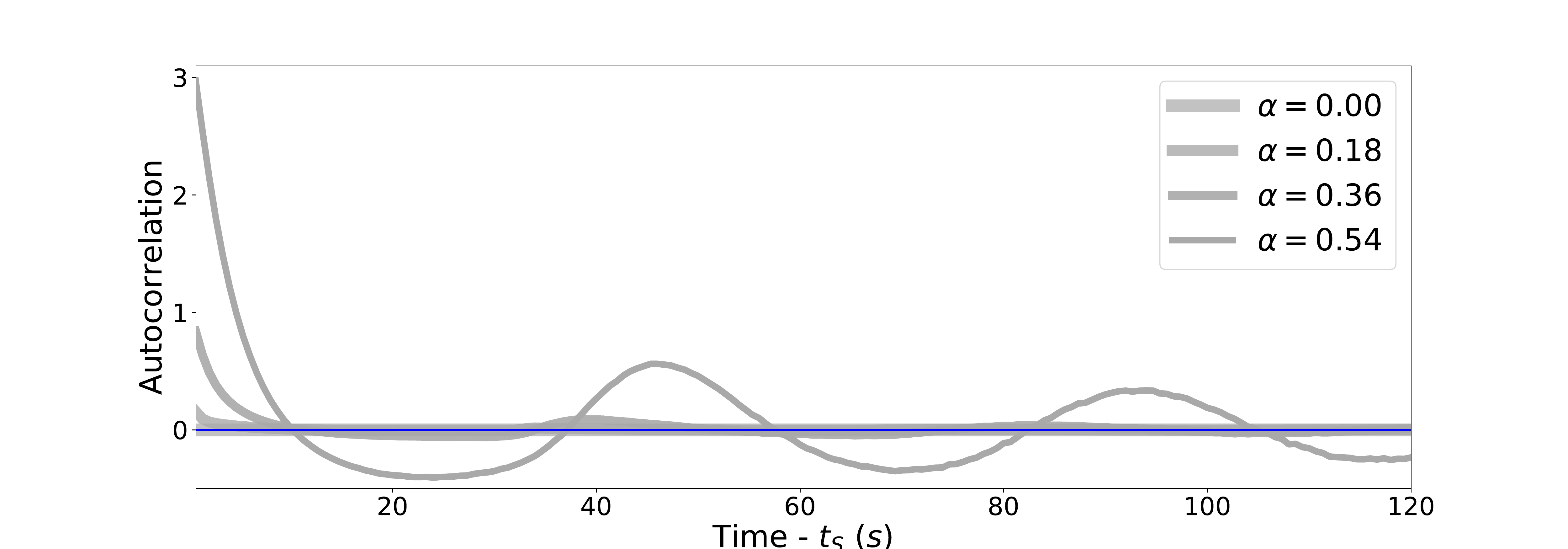}
	\caption{Mean temporal correlation function of the distance
          spacing in stationary states of the stochastic 
            Gipps model for
          different values of the noise amplitude $\alpha$. The
          oscillations of the autocorrelation arise when a noise is
          added and get more pronounced with increasing amplitude
          $\alpha$. The corresponding parameters are
          $a=1.7 \frac{m}{s^2}$, $\tau=\frac{2}{3}$~s,
          $v_{\rm max}=30\frac{m}{s}$, $b=3.0\frac{m}{s^2}$,
          $\hat{b}=2.9\frac{m}{s^2}$, $l=6.5$~m with $50$ cars on a
          ring of length $1000$~m.}
	\label{fig:3}
\end{figure}

Simulations are carried out in the deterministically stable regime
with periodic boundary conditions. A white noise with
amplitude $\alpha$ is added to the velocity.  The system can be
assumed to be in a stationary state after waiting for
$t_s= 2 \cdot 10^5$~s. In Fig.~\ref{fig:3} the autocorrelation of the
headway for different noise amplitudes is shown. The autocorrelation
begins to oscillate when a noise is added. The oscillations are
getting more pronounced when $\alpha$ is increased. The stable
homogeneous configuration is destabilized by the noise and oscillating
behaviour is observed in the stationary state.

Similar observations have been made in the social-force model as it
has been defined in \cite{sfm} when reduced to one dimension.
However, it is not reproducible in the first order OV model with a
white noise \cite{TordeuxS16}.
These findings indicate that the described mechanism is
not specific but rather generic for continuous models. It can be
reproduced if the model exhibits some inertia-like effect, such as a
(inertial) mass, a reaction time (incorporated as a delay or an update
time) or a relaxation time, as well as in first order models with a
correlated noise.

\section{Conclusion}

In continuous second order pedestrian models, stop-and-go
behaviour is produced if the inertia-related quantity is sufficiently
large.  However, this seems to lead generically to intrinsic
problems, like oscillations, which only can be avoided by choosing
unphysical parameters values. This has been already indicated by
earlier work \cite{Chraibietal,lakoba,Koester}.  The results presented
here strongly support this view and give new insights into the
mechanism behind this behavior.
Generically, the resulting effective inertia in these models is too
strong and drives the model into a regime of damped oscillations
instead of the overdamped region.
We have proposed an alternative approach based on a first order model
with a correlated noise that allows to overcome these problems while
choosing realistic parameter values. The underlying mechanism seems to
be generic and allows to overcome these problems for other models as
well.



%

\paragraph{Acknowledgments.}  Financial support by
  Deutsche Forschungsgemeinschaft (DFG) under grant SCHA~636/9-1 and
  Bonn-Cologne Graduate School of Physics and Astronomy (BCGS) and the
  German Excellence Initiative through the University of Cologne Forum
  "Classical and Quantum Dynamics of Interacting Particle Systems"\ is
  gratefully acknowledged.

%
%

\bibliographystyle{naturemag}
\bibliography{refs}

\end{document}